\documentclass[twocolumn,superscriptaddress]{revtex4}

\usepackage{graphicx}
\usepackage{amssymb}
\begin{document}

\title{Viscous fingering of miscible slices}

\author{A. De Wit}
\affiliation{Service de Chimie Physique and Center for Nonlinear
Phenomena and Complex Systems, Universit\'e Libre de Bruxelles, CP
231, 1050 Brussels, Belgium}

\author{Y. Bertho}
\affiliation{Service de Chimie Physique and Center for Nonlinear
Phenomena and Complex Systems, Universit\'e Libre de Bruxelles, CP
231, 1050 Brussels, Belgium} \affiliation{Microgravity Research
Center, Universit\'e Libre de Bruxelles, CP 165/62, 1050 Brussels,
Belgium}

\author{M. Martin}
\affiliation{Laboratoire PMMH-ESPCI (UMR 7636), 10 rue Vauquelin,
75\,231 Paris Cedex 05, France}

\begin{abstract}
Viscous fingering of a miscible high viscosity slice of fluid
displaced by a lower viscosity fluid is studied in porous media by
direct numerical simulations of Darcy's law coupled to the
evolution equation for the concentration of a solute controlling
the viscosity of miscible solutions. In contrast with fingering
between two semi-infinite regions, fingering of finite slices is a
transient phenomenon due to the decrease in time of the viscosity
ratio across the interface induced by fingering and dispersion
processes. We show that fingering contributes transiently to the
broadening of the peak in time by increasing its variance. A
quantitative analysis of the asymptotic contribution of fingering
to this variance is conducted as a function of the four relevant
parameters of the problem \emph{i.e.} the log-mobility ratio $R$,
the length of the slice $l$, the P\'eclet number $Pe$ and the
ratio between transverse and axial dispersion coefficients
$\varepsilon$. Relevance of the results is discussed in relation
with transport of viscous samples in chromatographic columns and
propagation of contaminants in porous media.
\end{abstract}

\maketitle

\section{Introduction}

Viscous fingering is an ubiquitous hydrodynamic instability that
occurs as soon as a fluid of given viscosity displaces another
more viscous one in a porous medium \cite{hom87}. As such, the
typical example usually presented for this instability is that of
oil recovery for which viscous fingering takes place when an
aqueous solution displaces more viscous oil in underground
reservoirs. This explains why numerous articles devoted to the
theoretical and experimental analysis of fingering phenomena have
appeared in the petroleum engineering community \cite{hom87}. For
what concerns the geometry, theoretical works typically focus on
analyzing the stability properties and nonlinear dynamics of an
interface between two {\it semi-infinite} domains of different
viscosity. In the same spirit, experimental works done either in
real porous media or in a model Hele-Shaw system (two parallel
plates separated by a thin gap width) consist in injecting {\it
continuously} a low viscous fluid into the medium initially filled
with the more viscous one. The attention is then focused on the
dynamics of the interface between the two regions. The instability
develops and the fingers grow continuously in time until the
displacing fluid has invaded the whole experimental system. As
long as the experiment runs (\emph{i.e.} until the displacing
fluid reaches the outlet), the instability develops. Dispersion of
one fluid into the other may lead to a slight stabilization in
time nevertheless this stabilization is usually negligible on the
time scale of the experiment and for high injection rates.

The situation is drastically different in other important
applications in which viscous fingering is observed, such as in
liquid chromatography or groundwater contamination. Liquid
chromatography is used to separate the chemical components of a
given sample by passing it through a porous medium. In some cases,
and typically in preparative or size exclusion chromatography, the
viscosity of the sample is significantly different than that of
the displacing fluid (the eluent). Displacement of the sample by
the eluent of different viscosity leads then to viscous fingering
of either the front or the rear interface of the sample slice,
leading to deformation of the initial planar interface. This
fingering is dramatic for the performance of the separation
technique as it contributes to peak broadening and distortions.
Such conclusions have been drawn by several authors that have
shown either experimentally
\cite{czo91,pla94,fer95,che97,bro98,fer96,nor96} or numerically
\cite{fer96,nor96} the influence of viscous fingering on peak
deformations.

In groundwater contamination and more generally soil
contamination, it is not rare that the spill's extent is finite
due to a contamination localized in space and/or time. If the
spill's fluid properties are different than that of water, and in
particular, if they have different viscosity and/or density
\cite{jia04,woo04}, fingering phenomena may influence the
spreading characteristics of the contaminated zone. For ecological
reasons, it is important then to quantify to what extent fingering
will enlarge the broadening in time of this polluted area.

Nonlinear simulations of fingering of finite samples have been
performed in the past by Tucker Norton, Fernandez \emph{et al.}
\cite{fer96,nor96} in the context of chromatographic applications,
by Christie \emph{et al.} in relation to ``Water-Alternate Gas"
(WAG) oil recovery techniques \cite{chr91} as well as by Zimmerman
\cite{zim04} and have shown the influence of fingering on the
deformation of the sample without however investigating the
asymptotic dynamics. Manickam and Homsy, in their theoretical
analysis of the stability and nonlinear dynamics of viscous
fingering of miscible displacements with nonmonotonic viscosity
profiles have further stressed the importance of reverse fingering
in the deformation of finite extent samples \cite{man93,man94}.
Their parametric study has focused on analyzing the influence of
the endpoint and maximum viscosities on the growth rate of the
mixing zone.

In this framework, the objectives of this article are twofold:
first, we analyze the nonlinear dynamics of viscous fingering of
miscible slices in typical analytical chromatographic and
groundwater contamination conditions in order to underline its
specificities and, second, we quantify the asymptotic contribution
of viscous fingering to the broadening of the output peaks as a
function of the important parameters of the problem. From a
numerical point of view, the only difference with regard to most
of the previous works devoted to viscous fingering
\cite{tan88,zim91,zim92} is the initial condition which is now a
sample of finite extent instead of the traditional interface
between two semi-infinite domains. As we show, this has an
important consequence: if the longitudinal extent of the slice is
small enough with regard to the length of the migration zone,
dispersion becomes of crucial importance as it leads to such a
dilution of the displaced sample into the bulk fluid before
reaching the measurement location that fingering just dies out. As
a consequence fingering is then only a transient phenomenon and
the output peak of the diluted sample may look Gaussian even if
its variance is larger than that of a pure diffusive dynamics
because of transient fingering. This explains why the importance
of fingering phenomena in chromatography and soil contamination
has been largely underestimated or ignored in the literature. We
perform here numerical simulations to compute the various moments
of the sample distribution as a function of time when fingering
takes place. This allows us to extract the contribution of viscous
fingering to the variance of the averaged concentration profile
and to understand how this contribution varies with the important
parameters of the problem which are the log-mobility ratio $R$
between the viscosity of the sample and that of the bulk fluid,
the P\'eclet number $Pe$, the dimensionless longitudinal extent
$l$ of the slice and the ratio $\varepsilon$ between the
transverse and longitudinal dispersion coefficients. The outline
of the article is the following. In Sec.~\ref{system}, we
introduce the model equations of the problem. Typical experimental
parameters for liquid chromatography and groundwater contamination
applications are discussed in Sec.~\ref{experimental}. The
characteristics of the fingering of a miscible slice are outlined
in Sec.~\ref{fingering}, while a discussion on the moments of
transversely averaged profiles is done in Sec.~\ref{moment}.
Eventually, a parametric study is conducted in
Sec.~\ref{parameter} before a discussion is made.

\section{Model system} \label{system}

\begin{figure}[t]
\includegraphics[width=8.5cm]{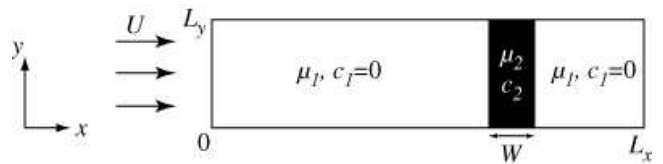}
\caption{Sketch of the system.} \label{Fig01}
\end{figure}

Our model system is a two-dimensional porous medium of length
$L_x$ and width $L_y$ (Fig.~\ref{Fig01}). A slice of fluid 2 of
length $W$ is injected in the porous medium initially filled with
carrier fluid 1. This fluid 2, which is a solution of a given
solute of concentration $c_2$ in the carrier, will be referred in
the following as the \emph{sample}.
This sample is displaced by the carrier fluid 1 in which the
solute concentration $c$ is equal to $c_1=0$. Assuming that the
viscosity of the medium is a function of the concentration $c$ and
that the flow is governed by Darcy's law, the evolution equations
for the system are then:
\begin{eqnarray}
\label{incomp}
\underline{\nabla}\cdot \underline{u} & = & 0,\\
\label{Darcy}
\underline{\nabla}p & = & -\frac{\mu(c)}{K}\underline{u},\\
\label{c} \frac{\partial c}{\partial t}+\underline{u} \cdot
\underline{\nabla}c & = & D_x \frac{\partial^2c}{\partial x^2}+D_y
\frac{\partial^2c}{\partial y^2},
\end{eqnarray}
where $\mu$ is the viscosity of the fluid, $K$ is the permeability
of the medium, $p$ is the pressure and $\underline{u}=(u,v)$ is
the two-dimensional velocity field. The displacing fluid is
injected in a uniform manner with a mean velocity $U$ along the
$x$ direction. $D_x, D_y$ are the dispersion coefficients along
the flow direction and perpendicular to it respectively. The
characteristic speed $U$ is used to define a characteristic length
$L_c=D_x/U$ and time $\tau_c=D_x/U^2$. We nondimensionalize space,
speed and time by $L_c, U$ and $\tau_c$ respectively. Pressure,
viscosity and concentration are scaled by $\mu_1D_x/K$, $\mu_1$
and $c_2$, where $\mu_1$ is the viscosity of the displacing fluid
and $c_2$ the initial concentration of the sample. The
dimensionless equations of the system become
\begin{eqnarray}
\underline{\nabla}\cdot \underline{u}&=&0,\\
\underline{\nabla}p&=&-\mu(c) \underline{u},\\
\frac{\partial c}{\partial t}+\underline{u} \cdot
\underline{\nabla}c &=& \frac{\partial^2c}{\partial x^2} +
\varepsilon \frac{\partial^2c}{\partial y^2},
\end{eqnarray}
where $\varepsilon=D_y/D_x$. If $\varepsilon=1$, dispersion is
isotropic while $\varepsilon \neq 1$ characterizes anisotropic
dispersion. Switching to a coordinate system moving with speed
$U$, \emph{i.e.} making the change of variables $x'=x-t, y'=y$,
$\underline{u}'=\underline{u}-\underline{i}_x$ with
$\underline{i}_x$ being the unit vector along $x$, we get, after
dropping the primes:
\begin{eqnarray}
\underline{\nabla}\cdot \underline{u} & = & 0,\\
\label{Dar}
\underline{\nabla}p & = & -\mu(c) (\underline{u}+\underline{i}_x), \\
\frac{\partial c}{\partial t}+\underline{u} \cdot
\underline{\nabla}c & = & \frac{\partial^2c}{\partial x^2} +
\varepsilon \frac{\partial^2c}{\partial y^2}.
\end{eqnarray}
We suppose here that the viscosity is an exponential function of
$c$ such as:
\begin{equation}
\mu(c)=e^{R c},
\end{equation}
where $R$ is the log-mobility ratio defined by
$R=ln(\mu_2/\mu_1)$, where $\mu_2$ is the viscosity of the sample
and, as said before, $\mu_1$ is the viscosity of the displacing
fluid (Fig.~\ref{Fig01}). If $R>0$, then we have a low viscosity
fluid displacing a high viscosity sample and the rear interface of
the sample will be unstable with regard to viscous fingering. If
$R<0$, then the sample is the less viscous fluid and the front
interface of the slice will then develop fingering. In our
simulations, we consider the $R>0$ situation.

Introducing the stream function $\psi$ such that $u=\partial
\psi/\partial y$ and $v=-\partial \psi/\partial x$, taking the
curl of Eq.~(\ref{Dar}), we get our final equations \cite{tan86}:
\begin{eqnarray}
&&\nabla^2 \psi = R \left ( \frac{\partial \psi}{\partial
x}\frac{\partial c}{\partial x} +\frac{\partial \psi}{\partial
y}\frac{\partial c}{\partial y} + \frac{\partial c}{\partial y}
\right ),\\
 &&\frac{\partial c}{\partial t}+\frac{\partial \psi}{\partial
y}\frac{\partial c}{\partial x} - \frac{\partial \psi}{\partial
x}\frac{\partial c}{\partial y} = \frac{\partial^2c}{\partial x^2}
+ \varepsilon \frac{\partial^2c}{\partial y^2}.
\end{eqnarray}

This model is numerically integrated using a pseudo-spectral code
introduced by Tan and Homsy \cite{tan88} and successfully
implemented for various numerical studies of fingering
\cite{dew99b,dew04}. The two-dimensional domain of integration is,
in dimensionless units, of size $Pe \times L$ where $Pe=UL_y/D_x$
is the dimensionless width which is nothing else than the P\'eclet
number of the problem, while $L=UL_x/D_x$. The dimensionless
length of the sample is $l=UW/D_x$. The initial condition
corresponds to a convectionless fluid ($\psi=0$ everywhere)
embedding a rectangular sample of concentration $c=1$ and of size
$Pe \times l$ in a $c=0$ background. The middle of the sample is
initially located at $x=2L/3$. In practice, for the simulations,
the initial condition corresponds to two back to back step
functions between $c=0$ and $c=1$ with an intermediate point where
$c=\frac{1}{2}+ A \cdot r$, $r$ being a random number between 0
and 1 and $A$ the amplitude of the noise (typically of the order
of $10^{-3}$). This noise is necessary to trigger the fingering
instability on reasonable computing time. If $A=0$, numerical
noise will ultimately seed the fingering instability but on a much
longer time scale. The boundary conditions are periodic in both
directions. This is quite standard along the transversal direction
$y$. This does not make any problem along the $x$-axis as $c=0$ at
both $x=0$ and $x=L$. The problem is controlled by four
dimensionless parameters: the log-mobility ratio $R$, the P\'eclet
number $Pe$, the initial length of the injected sample $ l$ and
the ratio between transverse and longitudinal dispersion
coefficients $\varepsilon$.

\section{Experimental values of parameters for two applications}
\label{experimental} In order to perform numerical simulations,
let us compute the order of magnitude of the main parameters
(P\'eclet number $Pe$, length of the sample $l$) for both a liquid
chromatography experiment and for the propagation of contaminants
in a porous medium (groundwater contamination).

\subsection{Chromatographic applications}
First of all, let us note that in most chromatographic
applications, heterogeneous chemistry (particularly adsorption and
desorption phenomena) is crucial to the separation process and
will undoubtedly affect possible fingering processes. We neglect
such physicochemical interactions in this first approach focusing
on the effect of viscous fingering on an unretained compound. A
typical chromatographic column has a diameter $d=4.6$\,mm, a
length $L_x=150$\,mm and consists of a porous medium packed with
porous particles, the total (intraparticle and interparticle)
porosity being equal to 0.7. The volume of the sample introduced
in the column is of the order of 20\,$\mu$l, injected at a flow
rate $Q\simeq 1$\,ml min$^{-1}$. The extent of the injected sample
is then $W\simeq 1.7\,10^{-3}$\,m and the speed of the flow
$U\simeq 1.4\,10^{-3}$\,m\,s$^{-1}$. The longitudinal and
transverse dispersion coefficients are typically
$D_x=1.43\,10^{-8}$\,m$^2$\,s$^{-1}$ \cite{kno99} and
$D_y=5.65\,10^{-10}$\,m$^2$\,s$^{-1}$ \cite{kno76}. These
parameters allow one to define a characteristic length $L_c=D_x/U$
and a characteristic time $\tau_c=D_x/U^2$. As a result, the
P\'eclet number $Pe=Ud/D_x$ is here nothing else than the
dimensionless diameter \emph{i.e.} $Pe=d/L_c\simeq 460$, while the
dimensionless longitudinal extent of the sample becomes
$l=W/L_c\simeq 170$. The dispersion ratio is equal to
$\varepsilon=D_y/D_x\simeq 0.04$. As a typical transit time from
inlet to outlet takes roughly $\tau=100$\,s, the dimensionless
time of a simulation should be of the order of
$T=\tau/\tau_c\simeq 15000$ to account for a realistic time to
characterize the properties of the output peaks.

\subsection{Soil contamination}
The effects of fluid viscosity and fluid density may be important
in controlling groundwater flow and solute transport processes.
Recently, a series of column experiments were conducted and
analyzed by Wood \emph{et al.} \cite{woo04} to provide some
insight into these questions. The experiments were performed in
fully saturated, homogeneous and isotropic sand columns (porosity
equals to 0.34 and $\varepsilon=1$) by injecting a 250\,ml pulse
of a known concentration solution at a flow rate
$Q=0.015$\,m$^3$/day. Their experimental setup consists of a
vertical pipe $L_x=0.91$\,m in length with a diameter of
$d=0.15$\,m. Assuming the medium to be homogeneous and the
dispersion coefficient $D$ as isotropic, a typical value for the
aquifer dispersion coefficient is $D=0.1$\,m$^2$/month \emph{i.e.}
$D\simeq 3.86\,10^{-8}$\,m$^2$\,s$^{-1}$ \cite{ser03}. In the same
spirit as above, we compute the P\'eclet number to be of the order
of $Pe\simeq 110$, while the dimensionless length of the sample is
$l\simeq 30$.

Based on these two examples, let us now investigate the properties
of fingering of finite slices for typical values of
parameters in the range computed above \emph{i.e.} $Pe \sim
100-500$, $\varepsilon \sim 0.04-1$, $l \sim 0-500$, while $R$ is
supposed to be of order one.

\section{Fingering of a finite width sample} \label{fingering}

\begin{figure}[t]
\includegraphics[width=8cm]{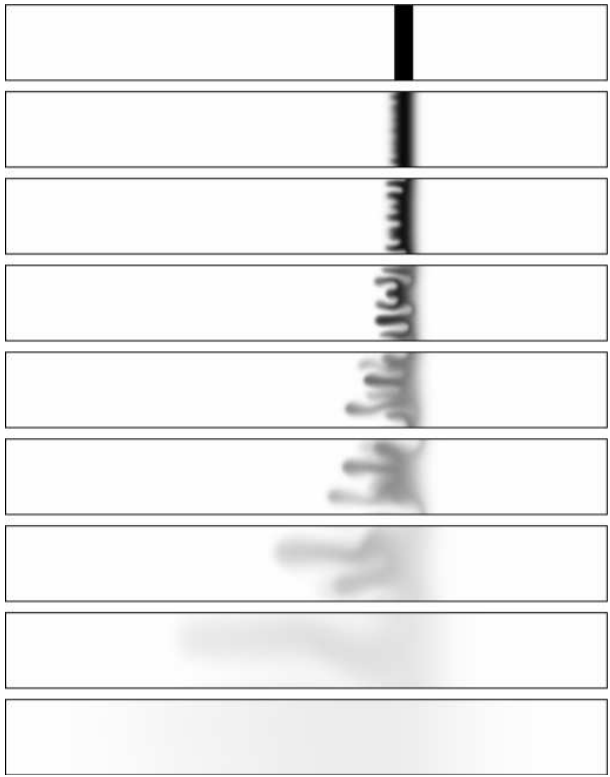}
\caption{Density plots of concentration at successive times in the
frame moving at the velocity $U$. From top to bottom: $t$=0, 500,
700, 1000, 1500, 2000, 5000, 15000 and 60000 ($Pe=512$, $l=128$,
$R=2$, $\varepsilon=1$).} \label{Fig02}
\end{figure}

Figure~\ref{Fig02} shows in a frame moving with the injection
velocity $U$ the typical viscous fingering of the rear interface
of a sample displaced from left to right by a less viscous fluid.
The system is shown at successive times using density plots of
concentration with black (resp. white) corresponding to $c=1$
(resp. $c=0$). While the front interface is stable, the back
interface develops fingers such that the center of gravity of the
sample is displaced in the course of time towards the back with
regard to its initial position. This dynamics results from the
fact that the stable zone acts as a barrier to finger propagation
in the flow direction leading therefore to reverse fingering. Such
a reverse fingering has been well characterized by Manickam and
Homsy in their numerical analysis of fingering of nonmonotonic
viscosity profiles \cite{man94}. After a while, dispersion comes
into play and dilutes the more viscous fluid into the bulk of the
displacing fluid. As the sample becomes more and more diluted, the
effective viscosity ratio decreases in time weakening the source
of the instability. Ultimately, dispersion becomes dominant and
the sample goes on diluting in the bulk without witnessing any
further fingering phenomenon. These successive steps can clearly
be observed on the transverse averaged profiles of concentration
defined as
\begin{equation}
\label{mean}
\bar{c}(x,t) = \frac{1}{Pe} \int_0^{Pe} c(x,y,t) dy.
\end{equation}
\begin{figure}[t]
\includegraphics[width=8cm]{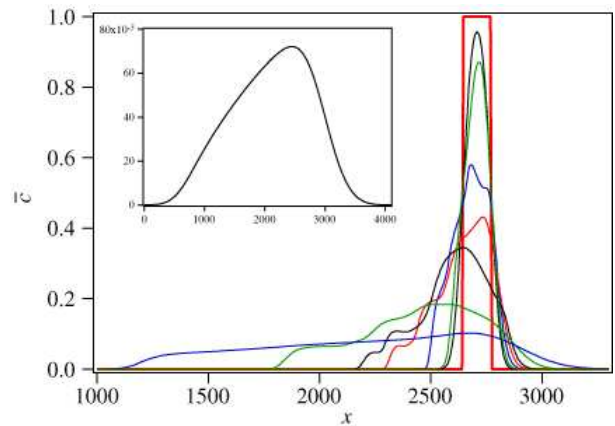}
\caption{Transverse average profiles of concentration at
successive times $t$=0, 500, 700, 1000, 1500, 2000, 5000 and
15000. Insert: transverse average profile of concentration at
$t=60000$ ($Pe=512$, $l=128$, $R=2$, $\varepsilon=1$).}
\label{Fig03}
\end{figure}
As seen on Fig.~\ref{Fig03}, we first start with two back to back
step functions defining an initial sample of extent $l$. During
the first stages of the injection, there is a first diffusive
regime quickly followed by the fingering of the rear interface
corresponding here to the left front. While the right (\emph{i.e.}
frontal) interface features the standard error function
characteristic of simple dispersion, the left one shows bumps
signaling the presence of fingering. Because the extent of the
sample is finite, dispersion and fingering contribute to the fact
that the maximum concentration becomes smaller than one,
effectively leading to a viscosity ratio between the sample and
the bulk that decreases in time. As a consequence, fingering dies
out and the transverse profile starts to follow a distorted
Gaussian shape. If one waits long enough, the asymmetry of the
bell shape diminishes which explains that output peaks in
chromatographic columns may look Gaussian even if fingering has
occurred during the first stages of the travel of the sample in
the column. As computed in the preceding section, a typical
dimensionless time of transit in a real chromatographic setup
corresponds to 15000 units of time. Figures~\ref{Fig02} and
\ref{Fig03} show that, after 15000 units of time, fingering is
disappearing for this specific set of typical values of
parameters, and that dispersion becomes again the dominant mode.
As chromatographic columns are generally opaque porous media, it
is therefore not astonishing that the presence of viscous
fingering has long been totally ignored until recent experimental
works which have visualized fingering by magnetic resonance or
optical imaging \cite{czo91,pla94,fer95,che97,bro98,fer96,nor96}.
Similarly, tracing of the spatial extent of a contaminant plume at
a distance far from the pollution site may lead to measurements of
Gaussian-type spreading even if fingering has occurred at early
times. The only influence of such fingering appears in the larger
variance of the sample than in the case of pure dispersion as we
show it next.

\section{Moments of the transverse averaged profile} \label{moment}

The averaged profiles of concentration $\overline{c}(x,t)$ allow
to compute the three first moments of the distribution: the
first moment $m$
\begin{equation}
\label{m1} m(t) = \frac{\int_0^{L} \bar{c}(x,t) x dx}{\int_0^{L}
\bar{c}(x,t) dx}
\end{equation}
is the position of the center of mass of the distribution as a
function of time. The second moment is the variance $\sigma^2$
\begin{equation}
\label{m2} \sigma^2(t) = \frac{\int_0^{L} \bar{c}(x,t) [x-m(t)]^2
dx}{\int_0^{L} \bar{c}(x,t) dx}
\end{equation}
giving information on the width of the distribution. Eventually,
we compute also the third moment, \emph{i.e.} the skewness
\begin{equation}
\label{m3} a(t) = \frac{\int_0^{L} \bar{c}(x,t)[x-m(t)]^3
dx}{\int_0^{L} \bar{c}(x,t) dx}
\end{equation}
that gives information concerning the asymmetry of the peak with
regard to its mean position.

The variance $\sigma^2$ is the sum of three contributions:
\begin{equation}
\label{ana}
\sigma^2(t) =\sigma^2_i+ \sigma^2_d+\sigma^2_f,
\end{equation}
where $\sigma^2_i=l^2/12$ is the variance due to the initial length
of the sample, $\sigma^2_d=2t$ is the contribution of dispersion in
dimensionless units and $\sigma^2_f$ is the contribution due to
the fingering phenomenon. If $R=0$, the displacing fluid and the
sample have the same viscosity and no fingering takes place.
Hence, in that case, $\sigma^2 = \sigma^2_i+\sigma^2_d = l^2/12+2
t$. We have checked that this result is recovered by numerical
simulations for $R=0$. The integrals in the computation of the
moments (\ref{m1})--(\ref{m3}) are evaluated numerically by using
Simpson's rule. The numerical result is very good if the spatial
discretization step $dx$ is small. Typically, we get the exact
result for $dx=1$. Unfortunately, $dx=1$ is a resolution too high
for fingering simulations especially if one wants to look at the
dynamics at very long times. As an example, previous simulations
on viscous fingering phenomena \cite{tan88,dew99b} were done
with larger $dx$ as typical dimensionless fingering wavelengths
are around 100 for $R=3$ for instance. Using typically $dx=4$
gives roughly 25 points per wavelength which is numerically
reasonable. For what concerns the variance, simulations with $R=0$
and $dx=4$ give the correct $\sigma^2_i$ at $t=0$ but a constant
shift appears so that $\sigma^2(t)-l^2/12-2 t =C$, with $C$ being
a constant of the order of $0.1\%$ of $l^2/12$. As we are mostly
interested in the rate of variation of $\sigma_{f}$, where
$\sigma_{f}$ is defined as
\begin{equation}
\sigma_{f} (t)=\sqrt{\sigma^2_{f}(t)} =\sqrt{\sigma^2(t)
-\sigma^2_i- \sigma^2_d},
\end{equation}
all simulations are done here with $dx=4$. The slight $C$ shift
does not affect the value of $\sigma_{f}$ as we have checked it
for decreasing values of $dx$.

\begin{figure}[t]
\includegraphics[width=8cm]{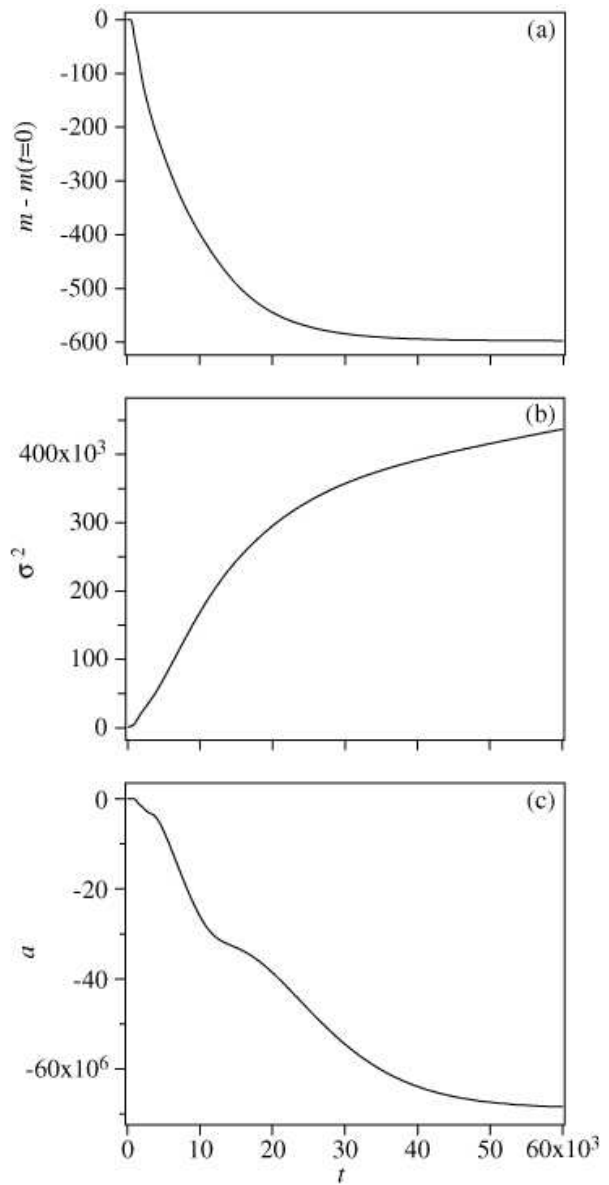}
\caption{First three moments of the distribution: (a)~Mean
position $m$ of the center of mass, (b)~Variance $\sigma^2$,
(c)~Skewness $a$ ($Pe=512$, $l=128$, $R=2$, $\varepsilon=1$).}
\label{Fig04}
\end{figure}
Figure~\ref{Fig04} shows the temporal evolution of the first three
moments \emph{i.e.} the deviation $m(t)-m(t=0)$ of the center of
mass in comparison to its initial location at $t=0$, the total
variance $\sigma^2(t)$ and the skewness $a(t)$ for the typical
example of Figs.~\ref{Fig02} and \ref{Fig03}. As fingering occurs
quicker than dispersion, the center of gravity of the sample
$m(t)$ is displaced towards the back (smaller $x$ values) because
of reverse fingering of the rear interface of the sample
[Fig.~\ref{Fig04}(a)]. Fingering contributes to the widening of
the peak and thus $\sigma^2(t)$ increases [Fig.~\ref{Fig04}(b)]
while the skewness $a(t)$ becomes non zero due to the asymmetry of
the fingering instability with regard to the middle of the sample
[Fig.~\ref{Fig04}(c)]. After a while, fingering dies out and the
first moment $m(t)$ saturates to a constant indicating that
dispersion becomes again the only important dynamical transport
mechanism. Note that the skewness $a$ is observed to revert back
towards 0 at very long times.

\begin{figure}[t]
\includegraphics[width=8cm]{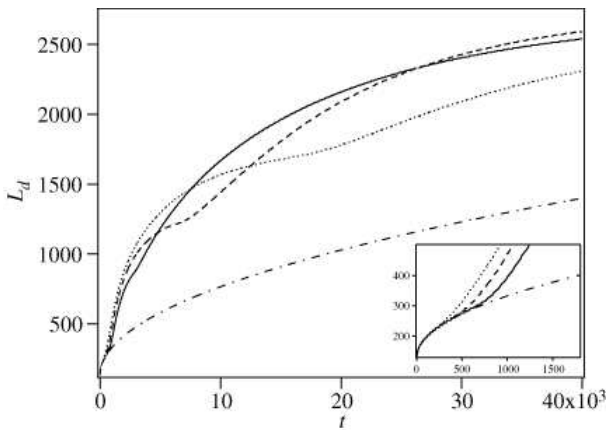}
\caption{Mixing zone $L_d$ as a function of time realized with the
same parameters ($Pe=512$, $l=128$, $R=2$, $\varepsilon=1$) and
three different values of the amplitude $A$ of the noise seeding
the initial condition: ($\cdots$)~$A=0.1$, (- -)~$A=0.01$,
(---)~$A=0.001$; (-- $\cdot$ --)~theoretical curve of a pure
diffusive behavior $L_d\propto \sqrt{t}$. Insert: zoom on the
first stages on the injection. The onset time $t^*$, corresponding
to the time at which the mixing zone departs from the pure
diffusive initial transient, is a decreasing function of $A$.}
\label{Fig05}
\end{figure}
Onset of fingering is also witnessed in the growth of the mixing
zone $L_d$ defined here as the interval in which $\bar{c}(x,t)>0.01$
(Fig.~\ref{Fig05}). An important thing to note is that, after a
diffusive transient, fingering appears on a characteristic time
scale $t^*$, defined as the time for which the mixing zone
temporal dependence departs from the pure diffusive regime.

As has already been discussed before \cite{zim91,dew04}, the
characteristics of the fingering onset time $t^*$ and of the
details of the nonlinear fingering regime are dependent on the
noise amplitude $A$. The higher the noise intensity $A$, the
quicker the onset of the instability (Insert in Fig.~\ref{Fig05}).
To get insight into the influence of the relevant physical
parameters of the problem, it is therefore necessary to fix the
amplitude of the noise to an arbitrary constant as this is not a
variable that is straightforwardly experimentally available. In
that respect, our results have here typically been obtained for a
noise of fixed $A=0.001$. The number of fingers appearing at early
times is related to the most unstable wavenumber of the band of
unstable modes, nevertheless the location and subsequent nonlinear
interaction of the fingers depend on the specific
realization of the random numbers series. As a consequence, it is
necessary to compute a set of realizations to get statistical
information on $\sigma_{f}$, the main quantity of interest here.
\begin{figure}[t]
\includegraphics[width=8cm]{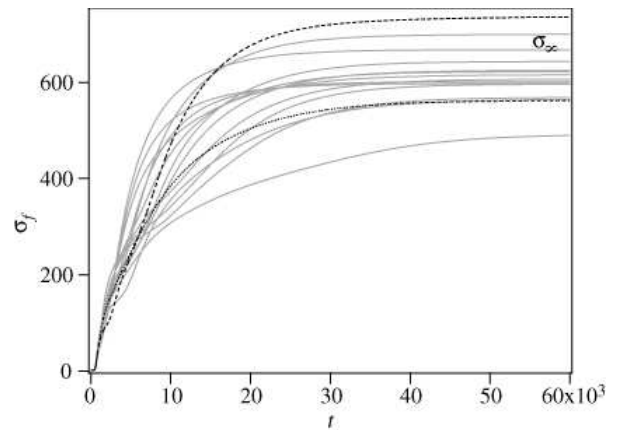}
\caption{$\sigma_{f}$ as a function of time for 15 numerical
simulations realized with the same values of parameters ($Pe=512$,
$l=128$, $R=2$, $\varepsilon=1$) but different noise $r$
realizations of identical amplitude $A=10^{-3}$. The dotted and
dashed curves correspond to the simulations of Figs.~\ref{Fig02}
 and \ref{Fig07} respectively.} \label{Fig06}
\end{figure}
Figure~\ref{Fig06} shows the temporal evolution of $\sigma_{f}$
for 15 different noise realizations of identical amplitude for
fixed values of the parameters $R, Pe, l$ and $\varepsilon$. As
can be seen, if fingering starts always at the same onset time
$t^*$ for fixed $A$, the contribution to the variance due to
fingering saturates to different asymptotic values
$\sigma_{\infty}$. This corresponds to slightly different
nonlinear interactions of the fingers as can be seen on
Figs.~\ref{Fig02} and \ref{Fig07} which show the temporal
evolution of the fingers for the respectively dotted and dashed
curves of Fig.~\ref{Fig06}. If the patterns observed are very
similar during the initial linear phase of viscous fingering, the
evolution of the fingers is slightly different in the nonlinear
regime, leading to different values of $\sigma_\infty$. In
particular, merging is observed in Fig.~\ref{Fig07} leading to the
fast development of one finger and, then, spreading of the stripe
of viscous fluid leading to a larger value of $\sigma_\infty$.
\begin{figure}[t]
\includegraphics[width=8cm]{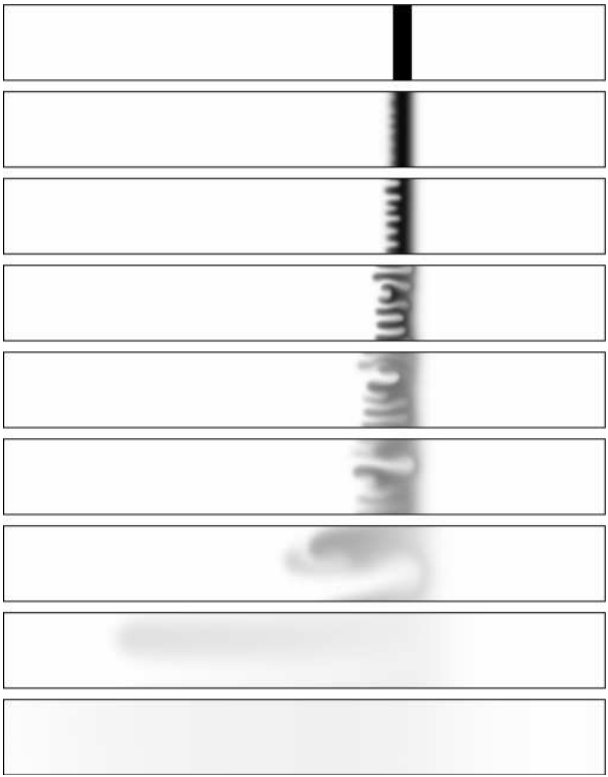}
\caption{Density plots of concentration for the same values of
parameters and same times as in Fig.~\ref{Fig02}, but a different
noise $r$ in the initial condition.} \label{Fig07}
\end{figure}

As a consequence, to understand the influence of fingering on the
broadening of finite slices, it is necessary to study the
parametric dependence of $<\sigma_{\infty}>$, the statistical
ensemble averaged asymptotic value of the fingering contribution
to the variance.

\section{Parameter study} \label{parameter}

The quantity $<\sigma_{\infty}>$ gives information on the
influence of viscous fingering on the broadening of finite
samples. In applications such as chromatography and dispersion of
contaminants in aquifers, such a broadening is undesirable and it
is therefore important to understand the optimal values of
parameters for which $<\sigma_{\infty}>$ is minimum given some
constraints. In that respect, let us first consider a porous
medium in which dispersion is isotropic ($\varepsilon=1$) and let
us analyze the subsequent influences of $l$, $R$ and $Pe$. The
anisotropic case ($\varepsilon \neq 1$) will eventually be
tackled. The mean value $<\sigma_{\infty}>$ is plotted for various
values of the parameters, the bar around this mean value spanning
the range of asymptotic data between the minimum and maximum
observed.

\subsection{Influence of the sample length $l$}

\begin{figure}[t]
\includegraphics[width=8cm]{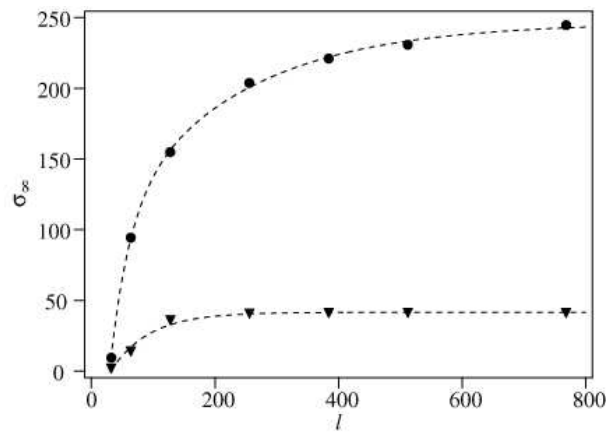}
\caption{Influence of the sample length $l$ on $\sigma_\infty$ for
($\blacktriangledown$)~$Pe=64$ and ($\bullet$)~$Pe=128$ ($R=2$,
$\varepsilon=1$).} \label{Fig08}
\end{figure}

The sample length $l$ has been measured to have practically no
influence on the onset time $t^*$ of the instability. Although
Nayfeh has shown that the stability of finite samples could be
affected if the two interfaces are close enough \cite{nay72}, we
note that, for the smallest value of sample length $l$ considered
here ($l=32$), the rear interface features the same initial
pattern as the one appearing on the interface between two
semi-infinite regions of different viscosities for a same random
sequence in the seeding noise. Our samples are thus here long
enough for the onset time $t^*$ to depend only on the amplitude
$A$ of the noise seeding the initial condition and not feel the
finite extent of the sample. The length $l$ influences
nevertheless the broadening of the peak and thus
$<\sigma_{\infty}>$ in particular for small $l$. The points
reported in Fig.~\ref{Fig08} for two different $Pe$ are obtained
for one realization and a same seeding noise $r$ in the initial
condition, leading to a typical value of $\sigma_{\infty}$. The
smaller the extent $l$ of the sample, the sooner the dilution of
the more viscous solution into the bulk of the eluent and thus the
less effective fingering. Above a given extent $l_c$,
$\sigma_{\infty}$ is found to saturate. At first sight, this might
appear counterintuitive as one could expect that, for longer
samples, fingering is maintained for a longer time thereby
enhancing the fingering contribution to the variance. A closer
inspection to the finger dynamics shows on the contrary that,
after a transient where several fingers appear and interact, only
one single finger remains (see Figs. 2 and 7). In the absence of
tip splitting, the stretching of the mixing zone becomes then
exclusively diffusive as already discussed previously by Zimmerman
and Homsy \cite{zim91,zim92}. This is clearly seen in Fig. 5 which
shows that the mixing length grows as $\sqrt{t}$ at long times
after a linear transient due to fingering. Once the asymptotic
diffusive regime is reached, the contribution of fingering to the
broadening of the peak dies out and $\sigma_{f}$ saturates to
$\sigma_{\infty}$. Above a given critical length $l_c$ of the
sample, the same asymptotic single finger growing diffusively is
reached before the left and right interfaces interact. Hence the
same value $\sigma_{\infty}$ is obtained for any $l>l_c$. Let us
note that the switch from the fingering to the diffusive dynamics
appears later in time when $Pe$ is increased. Indeed larger $Pe$
means more fingers that can interact for a longer time before the
diffusive regime becomes dominant. As a consequence, $l_c$ is an
increasing function of $Pe$ as can be seen on Fig.~\ref{Fig08}.
Further studies need to be done to understand the role of
$\varepsilon$ and of possible tip splitting occurring for large
$Pe$ on the existence and value of the critical length $l_c$.

The fact that the contribution of fingering to the broadening of
the peak saturates beyond a critical length of the sample has
important practical consequences for chromatography: if fingering
is unavoidable, one might as well load samples of long extent as
the contribution of fingering is saturating beyond a given $l_c$.
For long samples, the efficiency of the process depends then on
the competition between $\sigma^2_{\infty}$ and $l^2/12$, the
respective fingering and initial length contributions to the
peak's variance. We can thus predict that for $l_c < l <
\sigma_{\infty}/ \sqrt{12}$, the contribution of fingering is
constant and dominates the broadening while for $l >
\sigma_{\infty}/\sqrt{12}$, the initial sample length becomes the
key factor.

\subsection{Influence of the log-mobility ratio $R$}

\begin{figure}[t]
\includegraphics[width=8cm]{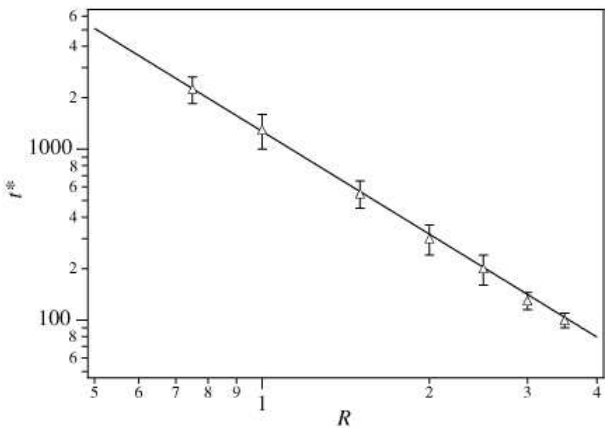}
\caption{Onset time $t^*$ of the instability for increasing values
of $R$ ($Pe=256$, $l=128$, $\varepsilon=1$). (---)~Best fit of the
experimental points: $t^*\propto R^{-2}$.} \label{Fig09}
\end{figure}
\begin{figure}[t]
\includegraphics[width=8cm]{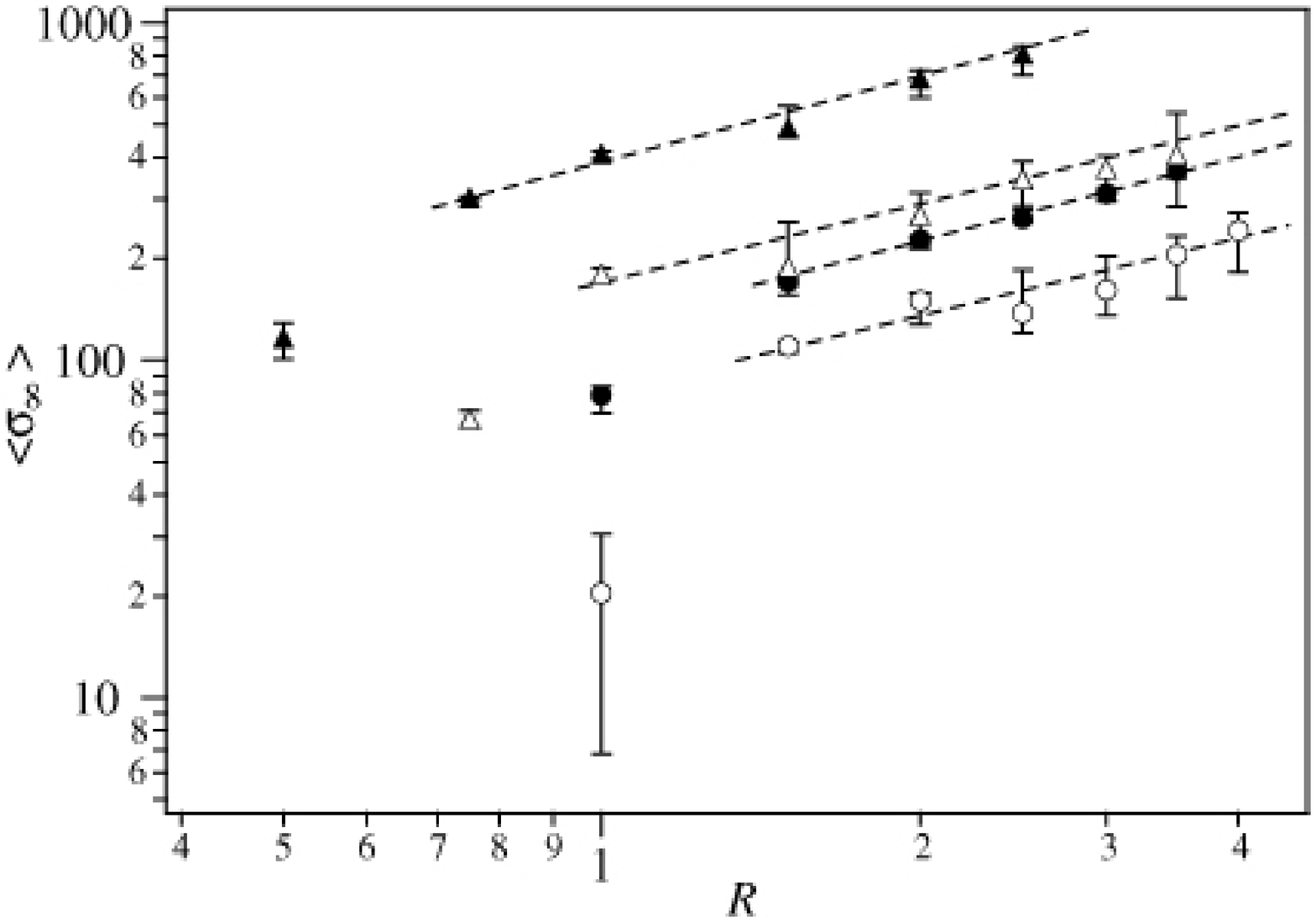}
\caption{Influence of the log-mobility ratio $R$ on
$<\sigma_{\infty}>$ for ($\circ$)~$Pe=128,l=128$,
($\bullet$)~$Pe=128,l=512$, ($\triangle$)~$Pe=256,l=128$,
($\blacktriangle$)~$Pe=256,l=512$ ($\varepsilon=1$).}
\label{Fig10}
\end{figure}

It is easy to foreseen that the larger $R$, the more important the
viscous fingering effect \cite{nor96}. First of all, linear
stability analysis of viscous fingering at the interface between
two semi-infinite domains \cite{tan86} predicts that the
characteristic growth time of the instability decreases as
$R^{-2}$. Although already influenced by the nonlinearities and
dependent on the amplitude of the noise, the onset time $t^*$
measured in our simulations shows the same trend
(Fig.~\ref{Fig09}). Note that, for very small values of $R$, the
onset time becomes very large which explains why, for samples of
low viscosity, fingering might not be observed during the transit
time across small chromatographic columns or on small scale
contamination zone. When $R$ is increased, the viscous fingering
contribution to peak broadening $<\sigma_{\infty}>$ is more
important (Fig.~\ref{Fig10}) with a linear dependence suggesting a
power law increase for larger $R$.

\subsection{Influence of the P\'eclet number $Pe$}

\begin{figure}[t]
\includegraphics[width=8cm]{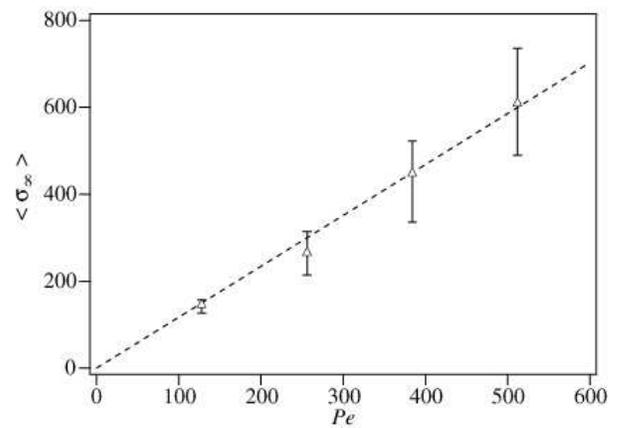}
\caption{Influence of the P\'eclet number $Pe$ on
$<\sigma_{\infty}>$ ($l=128$, $R=2$, $\varepsilon=1$).}
\label{Fig11}
\end{figure}
The P\'eclet number $Pe$ is typically experimentally increased for
a given geometry by increasing the injection flow rate $U$. As can
be seen on Fig.~\ref{Fig11}, $<\sigma_{\infty}>$ is found to
increase linearly with $Pe$. Fingering induced broadening can thus
be minimized by small carrier velocity $U$ as expected. However,
the exact influence of the carrier velocity $U$ is difficult to
trace because practically, a change in $U$ also modifies the
dispersion coefficients and hence the value of $\varepsilon$. In
our dimensionless variables, $U$ also enters into the
characteristic time and length corresponding respectively to
$D_x/U^2$ and $D_x/U$. The concrete influence of the carrier
velocity is thus more complicated to trace in reality. For a fixed
injection speed, the P\'eclet number can also be varied by
changing the width $L_y$ of the system. The linear dependence of
$<\sigma_{\infty}>$ on $Pe$ is then related to the fact that in a
wider domain, more fingers can remain in competition for a longer
time so that a more active fingering is maintained. This also
implies that $l_c$ is an increasing function of $Pe$. In
chromatographic applications, increasing the diameter of the
column (\emph{i.e.} increasing $L_y$ here in our model) is thus
expected to dramatically increase the influence of fingering in
broadening. This explains why fingering really becomes an issue
for wide contamination zones and in preparative chromatography
where columns of very large diameter (up to one meter) are
sometimes constructed.

\subsection{Influence of the ratio of dispersion coefficients $\varepsilon$}

\begin{figure}[t]
\includegraphics[width=8cm]{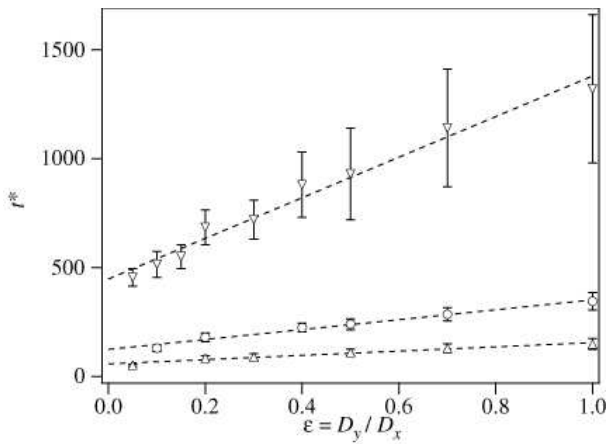}
\caption{Onset time $t^*$ of the instability for increasing values
of $\varepsilon$ for different values of the log-mobility ratio:
($\triangledown$)~$R=1$, ($\circ$)~$R=2$, ($\vartriangle$)~$R=3$
($Pe=128$, $l=128$).} \label{Fig12}
\end{figure}

\begin{figure}[t]
\includegraphics[width=8cm]{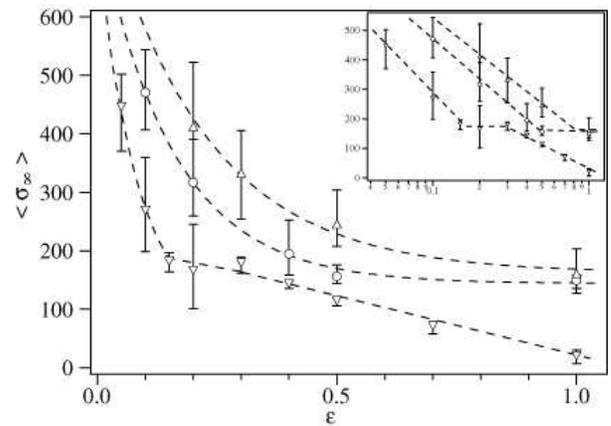}
\caption{Influence of the ratio between transverse and
longitudinal dispersion $\varepsilon$ on $<\sigma_{\infty}>$ for
($\triangledown$)~$R=1$, ($\circ$)~$R=2$, ($\triangle$)~$R=3$
($Pe=128$, $l=128$). Insert: same data on logarithmic scale for
$\varepsilon$.} \label{Fig13}
\end{figure}

Figure~\ref{Fig12} shows the influence of the ratio of dispersion
coefficients $\varepsilon=D_y/D_x$ on the onset time of the
instability. As expected from linear stability analysis
\cite{tan86}, decreasing $\varepsilon$ has a destabilizing effect
 as fingering appears then quicker. This is due to the
fact that small transverse dispersion inhibits the mixing of the
solutions and favors longitudinal growth of the fingers
allowing them to survive for a longer time. As a consequence, the
less viscous solution instead of being transversely homogeneous
invades the more viscous fluid preferably in the longitudinal
direction leading to larger mixing zones and hence larger
$<\sigma_{\infty}>$. Figure~\ref{Fig13} illustrates that decreasing
$\varepsilon$ has a dramatic effect on the broadening of the peak.
The insert shows the same graphics in logarithmic scale for
$\varepsilon$. $<\sigma_{\infty}>$ seems to vary as $\ln
(\varepsilon)$ at least for small values of $\varepsilon$. Peak
broadening due to fingering is therefore expected to be
particularly dramatic for chromatographic applications where
$\varepsilon\sim 0.04$.
\section{Conclusion}

Viscous fingering leads to a mixing between miscible fluids of
different viscosity. In the case of viscous slices of finite
extent, fingering is a transient phenomenon because the mixing of
the two fluids leads to an effective decrease of the log-mobility
ratio in time. Transient fingering can nevertheless play an
important role because it contributes to distortion and broadening
of the sample. In particular, we have shown that, even if the
spreading of the sample may look Gaussian at long times because
dispersion has again become the leading transport phenomenon, the
variance of the peak is larger than expected because of fingering
at early times. We have demonstrated this influence by numerical
simulations of viscous fingering of miscible finite slices
characterizing the onset time of the instability $t^*$ and the
contribution of fingering to the sample's variance. It is
important to note that quantitative comparison with experimental
data is difficult because the exact amplitude of the fingering
contribution to the temporal variation of the peak's variance
depends both on the amplitude and spatial realization of the noise
seeding the initial condition which varies from one experiment to
the other. In this respect, we have computed the ensemble averaged
asymptotic fingering contribution to the peak broadening as a
function of the four relevant parameters of the problem
\emph{i.e.} the initial length of the sample $l$, the initial
log-mobility ratio $R$, the P\'eclet number $Pe$ and the ratio
between transverse and longitudinal dispersion coefficients
$\varepsilon$. The broadening of the peaks due to fingering is
most important for large $R$ and $Pe$ but small $\varepsilon$
while it saturates above a given initial length $l$ of the sample.
In chromatographic columns for which $\varepsilon \sim 0.04$,
fingering is thus of crucial importance particularly in
preparative chromatography for which the large diameter of the
columns lead to large $Pe$ and the high concentration of the
samples usually implies large $R$. Similarly, for soil
contamination, fingering will be a major problem in the case of
stratified media such that $\varepsilon <1$. More work is now
needed to explore the generalization of this first approach to the
case where both viscosity and density variations as well as
heterogeneous chemistry may interplay as is usually the case in
the applications analyzed here.

\section{Acknowledgements}
We thank G.M. Homsy, P. Colinet, A. Vedernikov and B. Scheid for
fruitful discussions. Y. Bertho benefits from a postdoctoral
fellowship of the Universit\'e Libre de Bruxelles sponsored by the
Francqui Foundation which is gratefully acknowledged. A. De Wit
thanks also FRFC (Belgium) and the ``Communaut\'e fran\c caise de
Belgique - Actions de Recherches Concert\'ees" programme for
financial support.


\begin{thebibliography}{10}

\bibitem{hom87}
G.M. Homsy,
\newblock Viscous fingering in porous media,
\newblock {\em Ann. Rev. Fluid Mech.} {\bf 19}, 271 (1987).

\bibitem{bro98}
B.S. Broyles, R.A. Shalliker, D.E. Cherrak and G. Guiochon,
\newblock Visualization of viscous fingering in chromatographic columns,
\newblock {\em J. Chromatogr.} {\bf 822}, 173 (1998).

\bibitem{pla94}
L.D. Plante, P.M. Romano and E.J. Fernandez,
\newblock Viscous fingering visualized via magnetic resonance imaging,
\newblock {\em Chem.Eng. Sci.} {\bf 49}, 229 (1994).

\bibitem{fer95}
E.J. Fernandez, C.A. Grotegut, G.W. Braun, K.J. Kirschner, J.R. Staudaher, M.L. Dickson, and V.L. Fernandez,
\newblock The effects of permeability heterogeneity on miscible viscous fingering:
A three-dimensional magnetic resonance imaging analysis,
\newblock {\em Phys. Fluids} {\bf 7}, 468 (1995).

\bibitem{che97}
D. Cherrak, E. Guernet, P. Cardot, C. Herrenknecht and M. Czok,
\newblock Viscous fingering: a systematic study of viscosity effects in methanol-isopropanol systems,
\newblock {\em Chromatographia} {\bf 46}, 647 (1997).

\bibitem{czo91}
M. Czok, A. Katti and G. Guiochon,
\newblock Effect of sample viscosity in high-performance size-exclusion chromatography and its control,
\newblock {\em J. Chromatogr.} {\bf 550}, 705 (1991).

\bibitem{fer96}
E.J. Fernandez, T. Tucker Norton, W.C. Jung and J.G. Tsavalas,
\newblock A column design for reducing viscous fingering in size exclusion chromatography,
\newblock {\em Biotechnol. Prog.} {\bf 12}, 480 (1996).

\bibitem{nor96}
T. Tucker Norton and E.J. Fernandez,
\newblock Viscous fingering in size exclusion chromatography: insights from numerical simulation,
\newblock {\em Ind. Eng. Chem. Res.} {\bf 35}, 2460 (1996).

\bibitem{jia04}
C.Y. Jiao and H. Hotzl,
\newblock An experimental study of miscible displacements in porous media with variation of fluid density and viscosity,
\newblock {\em Trans. Porous Media} {\bf 54}, 125 (2004).

\bibitem{woo04}
M. Wood, C.T. Simmons and J.L. Hutson,
\newblock A breakthrough curve analysis of unstable density-driven flow and transport in homogeneous porous media,
\newblock {\em Water Resour. Res.} {\bf 40}, W03505 (2004).

\bibitem{chr91}
M.A. Christie, A.H. Muggeridge and J.J. Barley,
\newblock 3D simulation of viscous fingering and WAG schemes,
\newblock SPE 21238, presented at the 11th Symposium on Reservoir Simulation, Anaheim, California, February 17-20, 1991.

\bibitem{zim04}
W.B.J. Zimmerman, {\it Process Modelling and Simulation with Finite Element Methods} (World Scientific, Singapore, 2004).

\bibitem{man93}
O. Manickam and G.M. Homsy,
\newblock Stability of miscible displacements in porous media with nonmonotonic
viscosity profiles,
\newblock {\em Phys. Fluids A} {\bf 5}, 1357 (1993).

\bibitem{man94}
O. Manickam and G.M. Homsy,
\newblock Simulation of viscous fingering in miscible displacements
with nonmonotonic viscosity profiles,
\newblock {\em Phys. Fluids} {\bf 6}, 95 (1994).


\bibitem{tan88}
C.T. Tan and G.M. Homsy,
\newblock Simulation of nonlinear viscous fingering in miscible displacement,
\newblock {\em Phys. Fluids} {\bf 31}, 1330 (1988).

\bibitem{zim91}
W.B. Zimmerman and G.M. Homsy,
\newblock Nonlinear viscous fingering in miscible displacement with anisotropic dispersion,
\newblock {\em Phys. Fluids A} {\bf 3}, 1859 (1991).

\bibitem{zim92}
W.B. Zimmerman and G.M. Homsy,
\newblock Viscous fingering in miscible displacements : unification of effects of viscosity contrast, anisotropic dispersion, and velocity dependence of dispersion on nonlinear finger propagation,
\newblock {\em Phys. Fluids A} {\bf 4}, 2348 (1992).

\bibitem{tan86}
C.T. Tan and G.M. Homsy,
\newblock Stability of miscible displacements in porous media: rectilinear flow,
\newblock {\em Phys. Fluids} {\bf 29}, 3549 (1986).

\bibitem{dew99b}
A. De Wit and G.M. Homsy,
\newblock Nonlinear interactions of chemical reactions and viscous fingering in porous media,
\newblock {\em Phys. Fluids} {\bf 11}, 949 (1999).

\bibitem{dew04}
A. De Wit,
\newblock Miscible density fingering of chemical fronts in porous media: nonlinear simulations,
\newblock {\em Phys. Fluids} {\bf 16}, 163 (2004).

\bibitem{kno99}
J.H. Knox,
\newblock Band dispersion in chromatography - a new view of A-term dispersion,
\newblock {\em J. Chromatogr. A} {\bf 831}, 3 (1999).

\bibitem{kno76}
J.H. Knox, G.R. Laid and P.A. Raven,
\newblock Interaction of radial and axial dispersion in liquid chromatography in relation to the "infinite diameter effect",
\newblock {\em J. Chromatogr.} {\bf 122}, 129 (1976).

\bibitem{ser03}
S.E. Serrano,
\newblock Propagation of nonlinear reactive contaminants in porous media,
\newblock {\em Water Resour. Res.} {\bf 39}, 1228 (2003).

\bibitem{nay72}
A.H. Nayfeh,
\newblock Stability of liquid interfaces in porous media,
\newblock {\em Phys. Fluids} {\bf 15}, 1751 (1972).


\end{thebibliography}
\end{document}